\documentclass[referee]{jfm}
\usepackage{epsfig}
\usepackage[english]{babel}
\usepackage{natbib}
\usepackage{amsmath}
\usepackage{rotating}
\usepackage{upmath}
\usepackage{amssymb}
\usepackage{amsbsy}

\makeatletter
\gdef\@underjournal{%
  \vbox to 5.5\p@{\noindent
    \parbox[t]{4.5in}{\normalfont\indexsize{\itshape In
     preparation.}\\[2.5\p@]
      {\ \ }}%
  \vss}%
}
\makeatother

\newcommand{\Rey}{\mbox{\textit{Re}}}
\newcommand{\Rem}{\mbox{\textit{Rm}}}

\newcommand{\eps}{{\epsilon}}

\newcommand{\uvec}{{\mathbf{u}}}
\newcommand{\Bvec}{{\mathbf{B}}}
\newcommand{\Zvec}{{\mathbf{Z}}}

\newcommand{\Jvec}{{\mathbf{J}}}

\newcommand{\zvec}{{\mathbf{e}_z}}
\newcommand{\sigmam}{{\sigma_m}}

\newcommand\Ray{\mbox{\textit{Ra}}}  
\newcommand\Tay{\mbox{\textit{Ta}}}  

\bibpunct{(}{)}{;}{a}{}{,}

\begin{document}

\title{How do Dynamos Saturate?}
\author[F. Cattaneo and S. M Tobias]
{F\ls A\ls U\ls S\ls T\ls O \ns C\ls A\ls T\ls T\ls A\ls N\ls E\ls O$^1$
\and S\ls T\ls E\ls V\ls 
E\ls N\ns  M.\ns  T\ls O\ls B\ls I\ls A\ls S$^2$ 
}
\affiliation{$^1$ Department of Astronomy and Astrophysics and The Computation Institute, University of Chicago,
Chicago, IL 60637, USA \\[\affilskip]
$^2$ Department of Applied Mathematics, University of Leeds, Leeds LS2 9JT, UK
\\[\affilskip]}
\pubyear{2008}
\volume{???}
\pagerange{1-?}
\date{?? and in revised form ??}
\setcounter{page}{1}
\maketitle

\begin{abstract}
In order better to understand how dynamo systems saturate, we study the kinematic dynamo properties of velocity fields that arise from nonlinearly 
saturated dynamos. The technique is implemented by solving concurrently, in addition to the momentum equation, two induction equations, one for the actual magnetic field, and one for an 
independent passive field. We apply this technique to two illustrative examples: convectively driven turbulence, and turbulence represented by a shell model. In all 
cases we find that the velocity remains an efficient kinematic dynamo even after nonlinear saturation occurs. We discuss the implications to the process of dynamo 
saturation.%
\end{abstract}

\section{Introduction}
 
 The magnetization of a turbulent electrically conducting fluid is often conceptualised as
a two step process. Initially, a weak seed field is amplified by the turbulent motions.
During this kinematic phase, the field is assumed to be so weak  as to have no dynamical
effects on the turbulence. The fluid velocity is determined solely by the external, non
magnetic forces and, from the point of view of the magnetic field evolution, it can be
considered as prescribed. If, in this phase, the turbulent amplitude is stationary the
average behaviour of the magnetic field is either one of exponential growth or one of
exponential decay. However, it is now commonly believed that provided the magnetic Reynolds number, which is the non-dimensional measure of advection to diffusion, is high enough the magnetic
field will grow (Kazantsev 1968; Vainshtein \& Kichatinov 1986; Boldyrev and Cattaneo
2004). With the exponential growth of the magnetic field there will be a corresponding
exponential growth of the magnetic forces, which will eventually become comparable with
those driving the turbulence. In this second nonlinear phase the exponential growth of
the magnetic field will become saturated and the magneto-turbulence will settle down to
some stationary, well-defined level of magnetization.


By exploiting analogies with other fluid systems characterised by a growth phase followed
by a nonlinear saturation, it is possible to conceive of a number of saturation scenarios.
One possibility is that the nonlinear effects modify the background state and relax it
towards a condition of marginal stability. The classical example of a fluid system that
saturates in this way is thermal convection. Free convection occurs in thermally conducting
fluids when the mean temperature gradient exceeds that of an adiabatic atmosphere.
In the convective state, the effective
conductivity is greatly enhanced by the fluid motions relative to the collisional value
so that, for instance, the same energy flux can be transported by a shallower temperature
gradient. The saturated convective amplitude is then self-regulated to be the one that
allows the imposed energy flux to be transported along an  adiabatic gradient.
Another possibility is that the nonlinear interactions enhance the dissipation until it
balances the driving. The classical example of this type of saturation is hydrodynamic
turbulence. 
In the stationary regime the energy input
and viscous dissipation balance each other and the average state of the system is
determined by a distribution of energy at each scale such that a constant energy flux to
small scales can be maintained. For homogeneous, isotropic, incompressible turbulence
this corresponds to the celebrated Kolmogorov $k^{-5/3}$ spectrum.

In the case of a turbulent dynamo it is not likely that the saturation proceeds by relaxing the average state to a
marginal one.  First of all one should note that  turbulent dynamos can easily be greatly supercritical. In other words,
the magnetic Reynolds number, defined by $Rm=U\ell/\eta$, where $\eta$ is the magnetic diffusivity and $U$ and $\ell$ are
a characteristic velocity and length scales, respectively, can exceed the critical value for dynamo action (a few
hundreds for a turbulent fluid) by many orders of magnitude. Since the magnetic diffusivity is fixed, to relax the
average state to a marginal one would entail either a reduction of the velocity amplitude by several orders of magnitude
or a  reduction in characteristic scales by several orders of magnitude or a mixture of both.
Such dramatic changes are simply not observed. To be fair, in some cases, when dynamo saturation occurs the average
velocity does decrease to some extent, however, not by the huge amount required to bring $Rm$ close to its critical
value. A scenario in which driving and dissipation balance each other therefore seems more likely. In a turbulent dynamo the
driving is associated with the stretching of field lines, the dissipation by the destruction of magnetic flux by
reconnection; in a stationary state these two processes must balance. It can then be argued that the saturation process
consists of a modification of the turbulence such that this balance can be achieved. For example, in a turbulent flow, a
local measure of line stretching is afforded by the largest Lyapunov exponent; one could conceive that the nature of the
saturation is to reduce the average Lyapunov exponents. Indeed there are cases in which this reduction has been observed
(Cattaneo, Hughes  \& Kim 1996) but it does not seem to hold in general. Alternatively, one could argue that the smallest Lyapunov exponent
measures the local growth of gradients and hence the efficiency of dissipation; a balance could be reached by an average
decrease\footnote{Recall that the smallest Lyapunov exponent is negative} in the smallest Lyapunov exponent. However,
since in an incompressible fluid the sum of the Lyapunov exponents must be zero, a decrease in the negative exponent(s)
must be  accompanied by a corresponding increase in the postive one(s), which would lead to an increase in line
stretching efficiency.

The problem is that turbulent dynamo action is a subtle process and its saturation is correspondingly subtle. Two
difficulties can readily be identified: one is that, as noted by Kraichnan \& Nagarajan (1967), turbulent dynamos depends on the competition
between two exponential processes, stretching and enhanced dissipation; which is the dominant one cannot be ascertained
by dimensional arguments alone. The other is that, even if dynamo action is possible only if, on average,
stretching overcomes dissipation, it is far from obvious which average is the correct one. Should one use volume averages,
ensemble averages, or time averages along trajectories? Also, to what extent is it legitimate to swap the order between
taking exponentials and taking the averages?

In the present paper we address some of these issues. We study a related problem that can be precisely formulated and while providing some useful insight into the 
saturation process it is much simpler to analyze. We consider the turbulent velocity associated with a saturated dynamo and ask to what extent
this turbulent velocity acts as a kinematic dynamo. More specifically, we compare the turbulent velocity driven by large scale
forcing before and after the nonlinear saturation. We examine the ability of the resulting
velocity to amplify at an exponential rate a passive vector field that is not necessarily everywhere aligned with
the actual magnetic field, but whose evolution is determined by an induction equation with the same magnetic Reynolds
number as that for the magnetic field. Clearly, since both the passive field and the magnetic field obey the same linear equation,
if they are proportional to each other at one instant they will remain proportional forever. If, on the other hand, the two fields are
not everywhere aligned the passive field and magnetic field will have different evolutions. What we wish to compare are
their respective growth rates. In the kinematic regime if the magnetic field grows exponentially, any passive field will
eventually grow at the same rate. In this regime, one can regard the dynamo growth rate as an average property of the
velocity since it does not depend on which vector field it is applied to. It is not immediately obvious whether this will
continue to hold in the nonlinear regime. By definition, in the saturated state the growth rate of the magnetic field is
zero. But is this true for other vector fields, satisfying the induction equation? If the answer is yes, then the saturation property is an average property
in the same sense as above. If the answer is no, then the saturation property (by which we mean the property of neither growing nor decaying on average) is specific to some restricted class of
vector fields, one of which is the actual magnetic field, and does not hold in general.  

We apply this technique to two illustrative examples: convectively driven turbulence and MHD turbulence represented by a shell model.
Convective turbulence is a natural choice, it is known to be a effective dynamo, its properties can be carefully controlled
and it can be efficiently represented numerically. Shell models, on the other hand, provide an idealization of a turbulent flow
in which all degrees of freedom at a given wavenumber are represented by a single (complex) coefficient. 
They share some of the properties of the full systems, but lack their geoemetrical complexity.

\section{Convective Dynamos: Formulation}

We consider dynamo action driven by (Boussinesq) convection in a rotating plane layer. Using standard notations the evolution equations 
can be written as
\begin{eqnarray}
&
(\partial_t-\sigma\nabla^2)\uvec+\uvec\cdot\nabla\uvec +\sigma \Tay^{1/2}  \zvec \times \uvec =
-\nabla p + \Jvec\times\Bvec +\sigma\Ray\,\theta\zvec\;,
&\label{eq_mom}
\\
&
(\partial_t-\sigma/\sigmam\nabla^2)\Bvec+\uvec\cdot\nabla\Bvec =
\Bvec\cdot\nabla\uvec\;,
&\label{eq_ind}
\\
&
(\partial_t-\nabla^2)\theta+\uvec\cdot\nabla\theta = w\;,
&\label{eq_ene}
\\
&
\nabla\cdot\Bvec=\nabla\cdot\uvec=0\;,
&\label{eq_div}
\end{eqnarray}
where $\Jvec=\nabla\times\Bvec$ is the current density, $w$ is the vertical velocity, and $\theta$
denotes the 
temperature fluctuations relative to a linear background profile (e.g.\ Chandrasekhar 1961). In this 
non-dimensionalisation magnetic fields are measured in units of the Alfv\'en velocity. Four dimensionless numbers 
appear explicitly: the Rayleigh number $\Ray$, the Taylor number $\Tay$, 
and the kinetic and magnetic Prandtl numbers
$\sigma$ and $ \sigma_m$.

In the horizontal directions we assume that all fields are periodic with periodicity $\lambda$.
In the vertical we consider standard illustrative boundary conditions for the temperature,
velocity and magnetic field, namely
\begin{equation}
\theta=w=\partial_zu=\partial_zv=B_z=\partial_zB_x=\partial_zB_y=0 \quad {\rm at}~ z=0,1.
\label{eqn_bc1}
\end{equation}
We supplement (\ref{eq_mom}) -- (\ref{eq_div}) by an extra equation for the evolution of a (solenoidal) passive field, $\Zvec$, say, that obeys the same induction 
(\ref{eq_ind}) equation as $\Bvec$;

\begin{equation}
(\partial_t-\sigma/\sigmam\nabla^2)\Zvec+\uvec\cdot\nabla\Zvec =
\Zvec\cdot\nabla\uvec, \quad \nabla\cdot\Zvec=0,
\label{eq_pass}
\end{equation}
together with the same boundary conditions.

We solve equations (\ref{eq_mom}) -- (\ref{eq_pass}) numerically by standard
pseudo-spectral methods (see, for example, Cattaneo, Emonet \& Weiss (2003)).
For recent publications on convectively driven dynamos we note, for example, the work of Stellmach \& Hansen (2004) and  Cattaneo \& Hughes (2006). 

\section{Convective Dynamos: Results}

\begin{figure}
\centerline{\epsfig{file=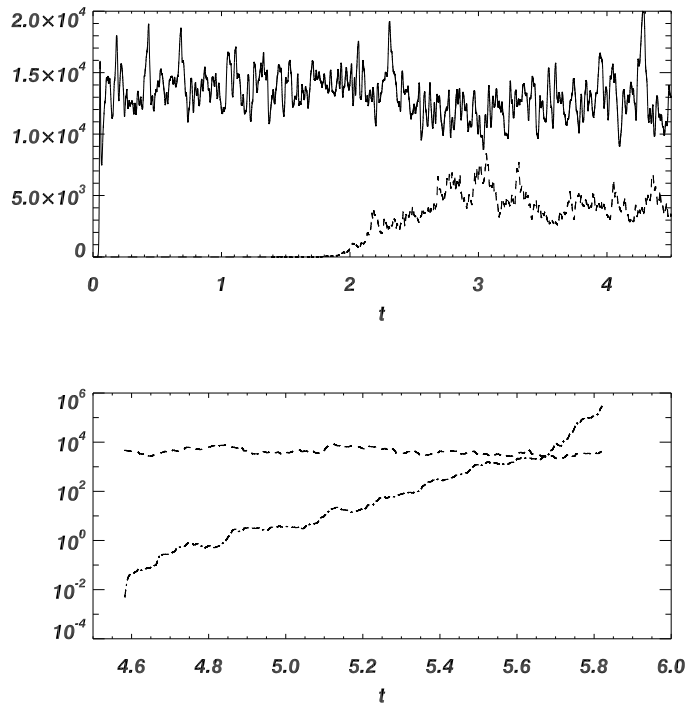,width=4truein}}
\caption{Convective dynamo timeseries. (a) Timeseries for the kinetic energy (solid line) and $5\times$ magnetic energy (dashed line) showing the dynamo 
evolution to a saturated state. (b) Timeseries for the magnetic energy (dashed line) and the energy of the passive field 
(dot-dashed line); note that this is logarithmic to show the exponential growth of the passive field energy.}
\label{nr_timeseries}
\end{figure}

\begin{figure}
\centerline{\epsfig{file=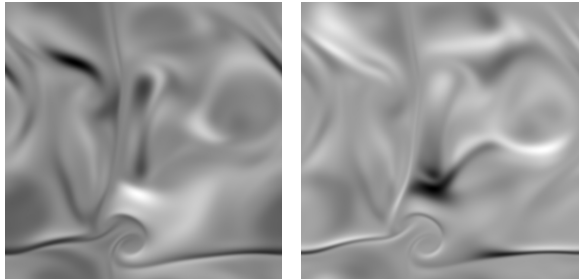,width=3truein}}
\caption{Density plots showing typical form of the magnetic field (left panel) and passive field (right panel) for the convective model. Shown are the 
$x$-components of both fields at a level $z \approx 0.1$.}
\label{nr_bxzx}
\end{figure}

The convective model 
equations (\ref{eq_mom}) -- (\ref{eq_pass}) were integrated for a wide range of parameters; with varying degrees of supercriticality (as measured by 
$Ra$) and rotation rates (as measured by $Ta$). For all cases, whether rotating or non-rotating, the results were qualitatively 
similar and so we focus here on describing the results for one representative case, with $Ta=0$ and $Ra=100,000$. The other 
parameters are set at $\sigma=1.0$, $\sigma_m=5.0$ and $\lambda=3.0$.

Figure~\ref{nr_timeseries}(a) shows the typical evolution of the convective dynamo system.
We first integrate the purely hydrodynamic system until a statistically steady convecting state, which consists of a 
number of moderately turbulent convective eddies interacting nonlinearly, is acheived. Once this hydrodynamic steady state is established (by 
$t=1$) a seed magnetic field $\Bvec_s$ is introduced --- in this stage the passive vector field $\Zvec$ 
remains zero. There follows a typical dynamo evolution for a convective flow. The seed magnetic field is amplified exponentially on 
an advective timescale (with a growth-rate $\sigma_{\Bvec} \approx 5.9$) and the system rapidly saturates in a turbulent 
magnetohydrodynamic state. Figure~\ref{nr_timeseries}(a) shows that the magnetic energy of the saturated state reaches approximately 
$10\%$ of the saturated kinetic energy, which is itself  reduced slightly (of the order of $10\%$) from its kinematic value. It is 
the dynamo properties of this saturated velocity field that is of interest.

To this end, we continue solving the equations for $\uvec$ and $\Bvec$, integrating the nonlinear saturated state forward in time. At $t 
\approx 4.75$ we introduce a random seed passive vector field ${\Zvec}_s$ into the linear equation~(\ref{eq_pass}). 
Figure~\ref{nr_timeseries}(b) shows the evolution of the saturated magnetic energy and the energy of the passive vector field as 
the calculation is then continued. As expected the saturated magnetic energy remains statistically steady throughout the evolution, 
but it is clear that the passive vector field is exponentially amplified by the saturated velocity field; the saturated 
velocity field {\it does indeed} act perfectly well as a dynamo! The growth-rate here is $\sigma_{{\Zvec}} \approx 5.3 < 
\sigma_{{\Bvec}}$
We stress here that if the initial passive vector field is not chosen to be random, but aligned with the saturated magnetic field 
--- i.e.\ ${\Zvec}_s = C \Bvec$, with $C$ a constant --- then ${\Zvec} = C \Bvec$ for all times in the subsequent 
evolution. 

Figure~\ref{nr_bxzx} compares the typical spatial form of the saturated magnetic field $\Bvec$ with that of the exponentially growing 
passive field $\Zvec$ in the form of density plots. It is clear that both fields have very similar spatial structures 
--- both are concentrated on small-scales. This is perhaps not surprising as they are advected by the same velocity  field at high 
magnetic Reynolds number, yet the passive field is growing exponentially whilst the magnetic field does not grow or decay on 
average. We stress again that these results are typical, and arise whether the system is moderately or highly turbulent; or rotating 
or non-rotating.

\section{Shell Model Dynamos: Formulation}

The results for the turbulent convective system described above seem counter-intuitive and there is the possibility that they may be model specific. We investigate 
this possibility by considering the simplest possible models of hydromagnetic dynamo action, namely shell models.

In the context of hydrodynamic turbulence, shell models have long been constructed with the aim of reproducing the spectral properties of turbulence within a low 
order model (see e.g. Gledzer 1973; Yamada \& Ohkitani 1987). More recently these shell models have been extended to include the effects of magnetic fields, both to 
examine dynamo action and magnetohydrodynamic turbulence (see e.g. Frick 1983; Plunian \& Stepanov 2007).
These models are constructed specifically to reproduce certain conservation laws that are inherent in the dissipation-free fluid and full magnetohydrodynamic systems 
and therefore to conserve ideal invariants. 

Here we consider the dynamics of a relatively simple magnetohydrodynamic shell model proposed by Frick \& Sokoloff (1998) (hereinafter FS98), in order to discuss the 
kinematic dynamo properties of a turbulent saturated dynamo. This local shell model conserves ideal invariants and can be shown to reproduce some dynamics of 
two-dimensional and three-dimensional MHD turbulence and dynamo action. As in the previous sections, we consider the evolution of the velocity field ($\uvec$), 
magnetic field ($\Bvec$) and the passive-vector field ($\Zvec$). Following FS98 we consider the dynamics on a range of spatial wavenumbers $k_n = k_0 \lambda^n$; $0 \le 
n \le n_{max}$ and consider the complex variable $U_n(t)$ as representative of all the modes of $\uvec$ in the shell with a wavenumber $k$ such that $k_n \le k \le 
k_{n+1}$. Similar representations for the magnetic fields and passive-vector field are given by the complex coefficients $B_n(t)$ and $Z_n(t)$.

The basic shell model as introduced in FS98 describes the evolution of the velocity coefficients ($U_n$) and the magnetic field coefficients ($B_n$) via the system
\begin{align}
\dot{U}_n+\Rey^{-1} k_n^2 U_n &= i k_n {\Bigg{\{}} \left( U^*_{n+1} U^*_{n+2} - B^*_{n+1} B^*_{n+2} \right)  
 - \dfrac{\eps}{2}
\left( U^*_{n-1} U^*_{n+1} - B^*_{n-1} B^*_{n+1} \right) \nonumber \\ 
 &- \dfrac{(1-\eps)}{4}
\left( U^*_{n-2} U^*_{n-1} - B^*_{n-2} B^*_{n-1} \right)
{\Bigg{\}}} + f_n
\label{ueqdef_shell}
\end{align}
\begin{align}
&\dot{B}_n+\Rem^{-1} k_n^2 B_n = i k_n \Bigg{\{} (1-\eps-\eps_m) \left( U^*_{n+1} B^*_{n+2} - B^*_{n+1} U^*_{n+2} \right) \nonumber \\
&+ \dfrac{\eps_m}{2}
\left( U^*_{n-1} B^*_{n+1} - B^*_{n-1} U^*_{n+1} \right)  
  + \dfrac{(1-\eps_m)}{4}
\left( U^*_{n-2} B^*_{n-1} - B^*_{n-2} U^*_{n-1} \right){\Bigg{\}}
\label{beqdef_shell}}
\end{align}

where $*$ represents the complex conjugate, $\Rey$ and $\Rem$ are the non-dimensional fluids and magnetic Reynolds numbers, $f_n$ is a random forcing acting on only 
a few shells near $n=0$. Here $\epsilon$ and $\epsilon_m$ are parameters, which are set at $\eps=1/2$, $\eps_m=1/3$ in order to conserve the relevant invariants 
(i.e.\ the total energy, the cross-helicity and the magnetic helicity) for non-dissipative three-dimensional dynamics (as in FS98). The only other parameter of the 
model is the spacing of the shells in wavenumber space ($\lambda$) which we set to $\lambda=\left( \sqrt{5}+1 \right)/2$, which is the minimum spacing allowed and is 
believed to lead to the most accurate results (see Plunian \& Stepanov 2007). 

This system of equations is able to describe regular dynamo action, and if the $B_n$ are set to zero then it reduces to the hydrodynamic GOY model. The dynamics of this 
system is described in detail in FS98. Here, as in the last two sections, we focus on achieving a saturated dynamo, where $U_n$ and $B_n$ reach statistically steady 
states, and examine the dynamo properties of the saturated velocity field $U_n$. This is achieved by simultaneously solving 
equations~(\ref{ueqdef_shell}-\ref{beqdef_shell}) together with the evolution equation for $Z_n$, which, of course, is identical to equation~(\ref{beqdef_shell}). It 
is the dynamics of this system of equations that will be investigated in the next section.

\section{Shell Model Dynamos: Results}

\begin{figure}
\centerline{\epsfig{file=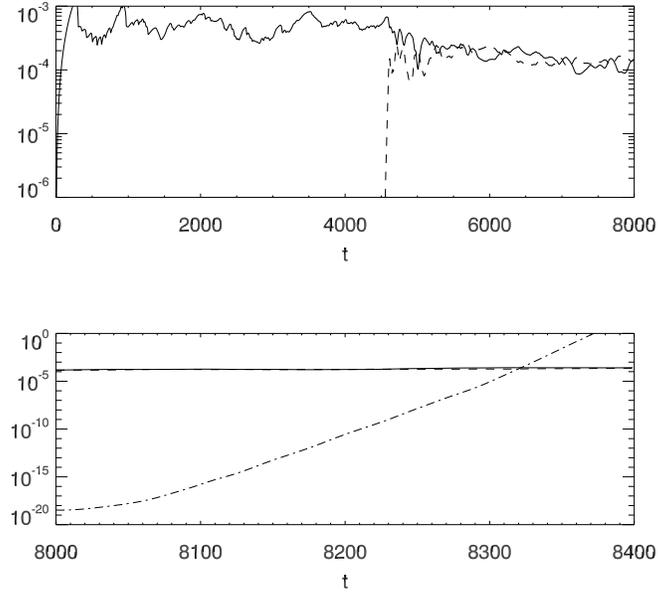,width=4truein}}
\caption{Shell model dynamo timeseries. (a) Timeseries for the kinetic energy (solid line) and  magnetic energy (dashed line) showing the dynamo 
evolution to a saturated state. (b) Timeseries for the kinetic energy (solid line), magnetic energy (dashed line) and the energy of the passive field 
(dot-dashed line).}
\label{shell_timeseries}
\end{figure}

\begin{figure}
\centerline{\epsfig{file=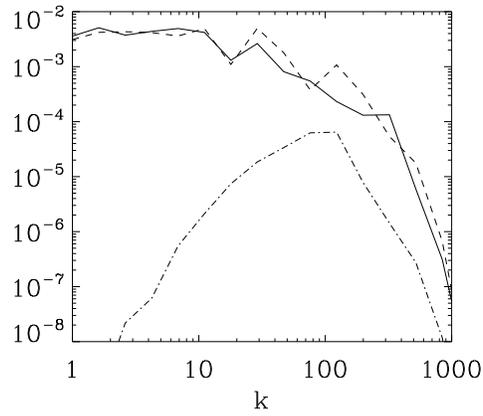,width=3.0truein}}
\caption{Spectra for the saturated velocity field (solid line), saturated magnetic field (dashed line) and kinematic passive field (dot-dashed).}
\label{shell_spectra}
\end{figure}

In this section, we repeat the procedure of \S3 for the shell model equations. Once again we integrate these 
equations for a wide variety of parameters and obtain qualitatively similar results each time. We present here a 
typical evolution, with $n_{max}=19$, corresponding to $k_{max} \approx 9349$. We fix $\Rey=\Rem=10^6$ and choose $f_n = 
10^{-4}\,(1+i)\, \delta_{n6}$ (so that steady forcing is applied at $n=6$). This choice of steady forcing leads to the driving of 
flows with a non-zero helicity; in shell models the helicity is defined as $H_u=\sum_{n} \, 0.5 (-1)^n k_n |U_n|^2$.

The hydrodynamic equations are integrated until a statistically steady state is achieved, as shown in 
Figure~\ref{shell_timeseries}(a). As for the convective model, this state has a complicated temporal evolution about a well-defined 
mean. Once the hydrodynamic solution has settled down, a small seed magnetic field is introduced and the dynamo evolution followed. 
Once again the seed magnetic field grows exponentially before saturating in a statistically steady MHD state (as shown in 
Figure~\ref{shell_timeseries}(a)). In this case the dynamo is very efficient and the magnetic energy saturates in equipartition 
with the kinetic energy. Figure~\ref{shell_spectra} shows the spectra for the velocity and magnetic field in the saturated state; 
it is clear from this that the shell-model has saturated the magnetic field so that it is in equipartition with the velocity field 
at each scale in the nonlinear regime. Interestingly the magnetic field does have significant power at small $k$ in the saturated 
state for this choice of parameters, having been localised to larger $k$ in the kinematic regime.

With the magnetic field and velocity field in a statistically steady saturated state, a weak passive seed field is added by 
setting $Z_n \ne 0$. The evolution of this passive field is compared with that of the saturated velocity and magnetic fields in 
Figure~\ref{shell_timeseries}(b). Once again the passive field grows exponentially, somewhat surprisingly with a growth-rate 
larger than the kinematic growth-rate for the magnetic field. Figure~\ref{shell_spectra} also compares the spectrum for the 
exponentially growing passive field at a representative time with that for the saturated magnetic field. It is clear that, in this case, the 
growing passive solution is more localised at high $k$ than the saturated magnetic field. We stress again that, although the 
details of the solutions are parameter dependent, the exponential amplification of the passive vector field remains a robust 
feature of the dynamics.

\section{Discussion}
%
%
%

In this paper we have addressed the issue of how dynamos saturate and have argued that this process is very subtle. In particular we have shown that in the saturated state the velocity remains a good kinematic dynamo for all passive vector fields that are not everywhere aligned with the magnetic field. Remarkably this holds both for full MHD systems, here we have analysed the specific case of convection, and for shell models. This implies that the dynamo does not saturate either by relaxing the system to a state close to marginality or by suppressing the chaotic stretching in the flow. Furthermore, because this result applies equally to full MHD and shell models, it suggests that the dynamo saturation relies on temporal rather than spatial correlations. 
It may be useful to draw an analogy between the dynamo system and that described by the product of $N$ random matrices with unit determinant. It is well-known that in general  the eigenvalues of such a product of random matrices grow exponentially with $N$. However, it {\it is} possible to construct sequences such that the eigenvalues of the product does not grow. 
The same matrices multiplied in a different order will however, in general, lead to exponential growth. Thus the stationarity of the eigenvalues is related to the specific order in which the matrices are multipled.
This specific ordering for the matrix problem is the equivalent of the temporal correlation for the dynamo problem. It is this subtlety that makes formulating a general theory for dynamo saturation so difficult.

This result also has some implications for the generation of large-scale fields. Thus far we have investigated the kinematic dynamo properties of a saturated velocity, with no particular distinction between large and small scale dynamo action. However we could focus this question on the case in which the saturated velocity is helical, or more generally lacks reflexional symmetry, and address the issue of large-scale dynamo action. What would be the evolution of a large-scale passive field advected by the saturated helical velocity? There are only two possibilities. Either the passive field is everywhere aligned with the actual magnetic field, in which case it will neither grow nor decay, or it is not aligned with the actual field, in which case it will quickly latch onto the fastest-growing `eigenfunction' and grow at the same rate as any other perturbation, small or large (Boldyrev \& Cattaneo 2004; Cattaneo  \& Hughes 2008). In general this fastest growing eigenfunction may be dominated by the large or small scales, but for the cases we considered the small scales dominated.

We conclude by summarising this work in what may sound like a truism, it is only the magnetic field that behaves like the magnetic field. Any other vector field that differs however little from the magnetic field will eventually behave in a qualitatively different way, despite solving the same evolution equation. In particular the averages of this other vector field will evolve differently from the averages of the real magnetic field;  one, for example,  will grow exponentially and the other will not.
It appears as though the evolution of any scale of magnetic field is intimately connected with the evolution of all the other scales. Thus, if one assumes that it is possible to derive an equation that correctly describes the evolution of the large-scale field in terms of average quantities, then one must also assume that this averaging procedure is sophisticated enough 
to take into account the chaotic dynamics of all the scales of the magnetic field --- including their temporal correlations.

\acknowledgements 
The authors would like to thank the Center for Magnetic Self Organisation and  The Leverhulme Trust, for support. 
This work was completed at the Dynamo Program at the Kavli Institute for Theoretical Physics, Santa Barbara, California (supported in part by the National Science Foundation under Grant No. PHY05-51164).

%
%
%

\end{document}